\documentstyle{article}
\title{\bf The Schwarzschild Black-Hole Pair}
\author{ Pedro F. Gonz\'alez-D\'{\i}az .\\
Instituto de Matem\'aticas y F\'{\i}sica Fundamental\\
Consejo Superior de Investigaciones Cientificas\\
Serrano 121, 28006 Madrid (SPAIN)\\
}
\date{March 5, 1995}
\begin{document}
\maketitle
\large
\setlength{\baselineskip}{0.5cm}
\vspace{3cm}

By allowing the lightcones to tip over according to the
conservation laws of an one-kink in static, Schwarzschild
metric, we show that there also exists an instanton which
represents production of pairs of chargeless, nonrotating
black holes with mass $M$, joined on an interior surface,
beyond the horizon at $r=M$. Evaluation of the thermal
properties of each of the black holes in a pair leads one
to check that each black hole is exactly the antiblack
hole to the other black hole in the pair. The instantonic
action has been calculated and seen to be smaller than
that corresponding to pair production by factors that
associate with the Bekenstein-Hawking entropy and a baby
universe entropy $2\pi^{2}M^{2}$. This suggests these
entropies to count numbers of internal states.

\vspace{5cm}


\noindent IMAFF-RC-04-95


\pagebreak

\section{\bf Introduction}
\setcounter{equation}{0}

The success of the thermodynamical analogy in black hole physics
allows us to hope that this analogy may be even deeper, and that
it is possible to develop a statistical-mechanical foundation
of black hole thermodynamics [1]. However, it is not quite clear
that black hole entropy may count the number of internal
degrees of freedom. These would describe different internal
states which may exist for the same values of the black hole
external parameters. It might be [2] that in the state of
thermal equilibrium the parameters for the internal degrees
of freedom will depend on the temperature of the system in
the universal way, thus cancelling all contributions which
depend on the particular properties and number of internal
fields.

Clearly, the most compelling argument in favour of the idea
that black hole entropy counts the number of internal states
has recently come from proposals of black hole pair creation [3-7].
By computing the exact action of the black hole pair instanton
it has been shown [4] that the action for the case of nonextreme
Reissner-Nordstrom black holes is smaller than that of the
corresponding pair creation rate by exactly a factor of
the black hole entropy $S_{BH}$; that is precisely what one
would expect if black holes had $e^{S_{BH}}$ internal states.
This is not the case nevertheless for extreme
Reissner-Nordstrom black holes where [5] the instanton action
exactly equals the rate of pair production, though this
apparent contradiction has been explained by the argument
that these black holes have zero entropy [6].

Apart from the fact that all these considerations are
based only on the leading order semiclassical approximation
and some higher order terms might be expected [7] to also
contribute importantly, the above analyses show two
limitations. First of all, they are restricted to deal
with Reissner-Nordstrom black holes. It would appear
to be of most interest if we could make a similar
analysis in Schwarzschild black holes, where there
is just an event horizon with unambiguous classical
localization and one external parameter, {\it i.e.}
the mass $M$. On the other hand, there seems to be
no clear idea about where exactly the internal states
may reside. They might be either inside the black hole
or on the horizon and, in the event the first possibility
applies, one may still wonder which region of the black
hole interior does form the geometrical domain for states,
so as exactly on what internal surface are identified the two black
holes in a pair while preserving the appropriate contribution
of the spacetime to the quantum construct.

An outcome to some of these limitations is attempted in
the present paper by considering the geometrical situation
that results (Section 2) from identifying a Schwarzschild
black hole with a spherically symmetric, four dimensional
gravitational topological defect which can move in spacetime
but cannot be removed without cutting [8]. As a consequence of
imposing invariance of the resulting kink number, we can
obtain a maximally extended Kruskal black hole metric which
is describable only by means of two coordinate patches. This
construct will represent two black holes, each in a different
universe, whose interior surfaces at $r=M$, rather than
the event horizons, are identified by the continuity of
the tipping over of lightcones. Thus, the two black holes
are joined by a nontraversible wormhole thinner than the
Einstein-Rosen bridge [9], leaving only a part of their
interior regions in the respective universe. By studying
the process of thermal emission of such Schwarzschild black
holes we have been able to show (Section 3) that they
form a black hole-antiblack hole pair, connected by a
Tolman-Hawking wormhole [10], even in the Lorentzian
case.
We have devised a method
to calculate (Section 4) the
Euclidean action of the corresponding instanton. It is seen
that this action is in fact smaller than that associated
with the semiclassical rate of pair production by factors
that correspond to the entropy of the black holes and to
an entropy of the resulting baby universe, $S_{U}=2\pi^{2}M^{2}$.
We conclude that both the entropy of the black holes and
the entropy of the baby universe would provide an actual count
of the internal states which are confined to reside,
respectively, in the region
between the event horizon and the surface at $r=M$, and
in the region beyond the latter surface, and can be
projected onto their respective enclosing surfaces.
We close up with
a summary of the results and some further comments. Unless
otherwise stated, we shall use natural units so that
$\hbar=c=G=1$, throughout the paper.

\section{\bf The Schawarzschild Kink Instanton}
\setcounter{equation}{0}

In this Section we shall derive an instantonic solution which
corresponds to the creation of a pair of Schwarzschild black
holes, joined at the internal surface $r=M$. This will be
accomplished by analytically continuing into the Euclidean
time the maximally extended Schwarzschild-Kruskal metric
kink. We first review the way along which this metric can
be obtained in a form that is suited for deriving the
instanton. We start with the usual, static
Lorentzian Schwarzschild metric
\begin{equation}
ds^{2}=-(1-\frac{2M}{r})dT^{2}+(1-\frac{2M}{r})^{-1}dr^{2}
+r^{2}d\Omega_{2}^{2},
\end{equation}
where $d\Omega_{2}^{2}$ is the metric on the unit 2-sphere.
Metric (2.1) is now transformed into the spherically symmetric
kink metric [11]
\begin{equation}
ds^{2}=\cos 2\alpha(-dt^{2}+dr^{2})-2\sin 2\alpha dtdr
+r^{2}d\Omega_{2}^{2},
\end{equation}
where $\alpha$ is the angle of tilt of the lightcones. We
ensure one kink to exist by requiring $\alpha$ to monotonously
increase from $0$ to $\pi$, starting with $\alpha(0)=0$.
Metric (2.2) can be transformed into (2.1) if we choose
\begin{equation}
\sin\alpha=\sqrt{\frac{M}{r}},
\end{equation}
and introduce the change of time coordinate $T=t+g(r)$, where
the function $g(r)$ must be chosen such that
\begin{equation}
\frac{dg}{dr}\equiv g'= \tan 2\alpha .
\end{equation}
$g'$ is singular at the event horizon where there appears
a geodesic incompleteness. Since $\sin\alpha$ cannot exceed
unity it follows that $\infty\geq r\geq M$, such that
$\alpha$ varies from $0$ to $\frac{\pi}{2}$ only. In order
to describe a complete one-kink one need therefore a second
coordinate patch for $\frac{\pi}{2}\leq\alpha\leq\pi$. The
need for such an additional coordinate patch can be better
seen by introducing a new time coordinate $\bar{t}=t+h(r)$,
where the new function $h(r)$ is defined so that
\begin{equation}
\frac{dh}{dr}\equiv h'=\tan 2\alpha-\frac{k_{1}}{\cos 2\alpha},
\end{equation}
with $k_{1}=\pm 1$. Then $k_{1}=+1$ will correspond to the
first coordinate patch and $k_{1}=-1$ to the second one.
This time re-definition transforms metric (2.2) into the
standard kink metric [11]
\begin{equation}
ds^{2}=\cos 2\alpha d\bar{t}^{2}-2k_{1}d\bar{t}dr
+r^{2}d\Omega_{2}^{2}.
\end{equation}
In metric (2.6) the existence of two coordinate patches
corresponding to the two possible values of $k_{1}$ is
clearly manifested.

However, geodesic incompleteness at $r=2M$ is still present
in the two coordinate patches of metric (2.6). This
geodesic incompleteness can be removed by using the Kruskal
method. Thus, let us introduce the general metric
\begin{equation}
ds^{2}=-2F(U,V)dUdV+r^{2}d\Omega_{2}^{2},
\end{equation}
where
\begin{equation}
U=\mp e^{\gamma\bar{t}}e^{2\gamma k_{1}\int_{\infty/M}^{r}\frac{dr}{\cos
2\alpha}}
\end{equation}
\begin{equation}
V=\mp\frac{1}{2\gamma M}e^{-\gamma\bar{t}}
\end{equation}
and
\begin{equation}
F=\frac{4M\cos 2\alpha}{\gamma}e^{-2\gamma
k_{1}\int_{\infty/M}^{r}\frac{dr}{\cos 2\alpha}}.
\end{equation}
The arbitrary parameter $\gamma$ will be chosen such that the
resulting metric will not show any singularity other than the
curvature singularity at $r=0$, and the lower integration limit
$\infty/M$ refers to the choices $r=\infty$ and $r=M$, depending
on whether the case $k_{1}=+1$ or the case $k_{1}=-1$ is being
considered. Let us then evaluate the integral appearing in the
exponent of (2.8) and (2.10)
\[\int_{\infty/M}^{r}\frac{dr}{\cos 2\alpha}
= 2M(\frac{1}{2}\csc^{2}\alpha -
\ln(\frac{\sin^{2}\alpha}{2\sin^{2}\alpha-1})),\]
where we have not explicited any real constants coming from lower
integration limits. Such constants can always be absorbed in a
normalizing factor in the definition of $U$ and $F$.

We avoid the unphysical singularity at $r=2M$ by setting $\gamma^{-1}
=4Mk_{1}$. Hence,
\begin{equation}
U=\mp e^{\frac{\bar{t}}{4Mk_{1}}}e^{\frac{r}{2M}}(\frac{2M-r}{M})
\end{equation}
\begin{equation}
V=\mp 2k_{1}e^{-\frac{\bar{t}}{4Mk_{1}}}
\end{equation}
and
\begin{equation}
ds^{2}=-\frac{32M^{3}k_{1}}{r}e^{-\frac{r}{2M}}dUdV+r^{2}d\Omega_{2}^{2}.
\end{equation}
The use of (2.11) and (2.12) transforms (2.13) into (2.6). Metric
(2.13) is the same as the kinkless extension of the Schwarzschild
metric unless by the sign ambiguity parameter $k_{1}$ which
distinguishes the two coordinate patches.

In order to look for the instanton that corresponds to metric
(2.13), we should start with the Schwarzschild instanton
derived from (2.1) by the replacement $T\rightarrow i\tilde{T}$.
Then, $\tilde{T}=\tilde{t}+\tilde{g}(r)$, $\tilde{\bar{t}}
=\tilde{t}+\tilde{h}(r)$, so that $g'\rightarrow i\tilde{g}'$,
$h'\rightarrow i\tilde{h}'$ and metrics (2.2) and (2.6)
respectively become
\begin{equation}
d\tilde{s}^{2}=\cos 2\alpha(d\tilde{t}^{2}+dr^{2})
-2i\sin 2\alpha drd\tilde{t}+r^{2}d\Omega_{2}^{2}
\end{equation}
\begin{equation}
d\tilde{s}^{2}=\cos 2\alpha d\tilde{\bar{t}}^{2}
-2ik_{1}drd\tilde{\bar{t}}+r^{2}d\Omega_{2}^{2}.
\end{equation}
The Kruskal extension of (2.15) will lead finally to the metric
\begin{equation}
d\tilde{s}^{2}=+\frac{32M^{3}k_{1}}{r}e^{-\frac{r}{2M}}d\tilde{U}d\tilde{V}
+r^{2}d\Omega_{2}^{2},
\end{equation}
where $\tilde{U}=U$ and $\tilde{V}=V$. Therefore, the instantonic
metric (2.16) is the same as the Lorentzian metric (2.13), unless
by the sign of the first term in the rhs.
In Eqn. (2.5) we chose the sign minus in front of $k_{1}$ in the
second term of the rhs. A corresponding plus sign had exchanged
the roles played by the two coordinate patches. This is exactly
the sole effect we obtain by Euclideanizing metric (2.13).
In what follows we shall discuss the meaning of the instanton
(2.16) for which
\begin{equation}
\tilde{U}\tilde{V}=UV=2k_{1}e^{\frac{r}{2M}}(\frac{2M-r}{M}).
\end{equation}

The Kruskal diagrams for the two coordinate patches are given
in Fig. 1, in the case of the Lorentzian metric. The corresponding
diagrams for the Euclidean metric can be simply obtained from
those of Fig.1 by changing the sign of $k_{1}$ both in each
patch and in the original regions ($I_{k_{1}},II_{k_{1}}$) and the
new regions created in the Kruskal extension ($III_{k_{1}},IV_{k_{1}}$).
We first note that the curvature singularity at $r=0$ is avoided
because of the continuity of the angle of tilt $\alpha$ on
$\alpha=\frac{\pi}{2}$. This in turn implies identification
of hypersurfaces $r=M$ of the two patches, both for original
and new regions, and Lorentzian and Euclidean metrics. These
identifications amount to the existence of bridges connecting
the two coordinate patches.

In order to analyse physical processes taking place in the
diagrams it is instructive to consider paths followed by
null geodesics on them. For the geodesic labelled
$a_{1}a_{2}a_{3}$, the segment $a_{1}$, which starts at
$r=\infty$, crosses the $V$ axis
on the event horizon $r=2M$, and then passes from original
region $I_{+}$ to $II_{+}$. It continues as segment $a_{2}$
passing into original region $I_{-}$ of the second patch,
to cross after $t=-\infty$ ($U$ axis) into the new region
$III_{-}$, to end up at $r=\infty$ on this region. The
whole process corresponds to the transition from an
asymptotically flat region in a coordinate patch to another
asymptotically flat region in the other coordinate patch,
both in Lorentzian spacetime. It takes place through a
bridge (wormhole) with size $r=M$, in a similar fashion to
that occurs in the Einstein-Rosen bridge [9]. The main an essential
difference is that in the present case the connection happens
at $r=M$, rather than $r=2M$.

Complete transitions with null geodesics in our Lorentzian spacetime
cannot occur however since the crossing of the horizon in patch
$k_{1}=-1$ will take an infinite Lorentzian time $t$ in the
process considered for geodesic $a_{1}a_{2}a_{3}$. This
leaves any null ray trapped inside the black hole of patch
$k_{1}=-1$ whose asymptotic region therefore will never be
reached. This difficulty is no longer present in the Euclidean
metric where a true instantonic tunnelling connecting the
asymptotic regions of the two patches makes the transition
possible. To be sure, in the Euclidean case paths would also cross
the horizon $\tilde{V}=0$, but this occurs now at an infinite
{\it imaginary} time, outside the lightcones. Passage through
the bridge can thus be regarded as a quantum tunnelling between two
distinct universes with the branching off of a baby universe
the size $M$. This can better be illustrated by using, instead of
$\bar{t}$, the new
coordinates $\tau=f(\chi)+t$, $r=a(\tau)\sin\chi$,
with $0\leq\chi\leq\pi$,
and restricting to the surface $r=M$ ({\it i.e.}
$\alpha=\frac{\pi}{2}$), with
\[\frac{df}{d\chi}=ak\sin\chi=kM, \; \; \;
\frac{da}{d\tau}=\frac{k}{\cos\chi},\]
where again $k=\pm 1$ for the two coordinate patches. Metric (2.2)
on $r=M$ can then be rewritten
\begin{equation}
ds_{W}^{2}=(1-\frac{M^{2}}{a^{2}})^{-1}d\tau^{2}+a^{2}d\Omega_{3}^{2}
=a^{2}\left(d\eta^{2}+d\Omega_{3}^{2} \right)
\end{equation}
\begin{equation}
a=(M^{2}+t^{2})^{\frac{1}{2}}=Me^{\eta}, \; \; \; f=kM\chi + Const.,
\end{equation}
in which $d\Omega_{3}^{2}=d\chi^{2}+\sin^{2}\chi d\Omega_{2}^{2}$
is the metric on the unit 3-sphere. We have
thus attained the metric of the Tolman-Hawking wormhole [10], with
the scale factor given in terms of the Lorentzian time $t$ or
$\eta=\int\frac{d\tau}{t}$.
In the Euclidean picture, from (2.14) on $r=M$ and
$i\tilde{\tau}=\tilde{f}(\chi)+\tilde{t}$, $r=a(\tilde{\tau})\sin\chi$,
instead of $\tilde{\bar{t}}$, we arrive at
\begin{equation}
ds_{U}^{2}=(1-\frac{M^{2}}{a^{2}})^{-1}d\tilde{\tau}^{2}
+a^{2}d\Omega_{3}^{2}=a^{2}\left(d\tilde{\eta}^{2}+d\Omega_{3}^{2}\right)
\end{equation}
\begin{equation}
a= (M^{2}-\tilde{t}^{2})^{\frac{1}{2}}=Me^{i\tilde{\eta}},
\; \; \; \tilde{f}=ikM\chi + Const.,
\end{equation}
which is the metric of the Tolman (baby) universe [10], now with the
scale factor expressed in terms
of the Euclidean time $\tilde{t}$ or $\tilde{\eta}=
\int\frac{d\tilde{\tau}}{\tilde{t}}$.

It appears then that a connection between black holes and wormholes
can be achieved if we let lightcones of a single black hole
spacetime to tip over from $\alpha=0$ to $\alpha=\frac{\pi}{2}$
in a given coordinate patch. Then, conservation of the resulting
kink number would imply the emergence of a second black hole in
a different coordinate patch whose lightcones would tip over
between $\frac{\pi}{2}$ and $\pi$. On the internal surface
$\alpha=\frac{\pi}{2}$ the two balck holes are
connected through a Tolman-Hawking wormhole (in the Lorentzian
case) or a baby universe (in the instantonic case).

It is known that the complete kinkless Schwarzschild-Kruskal
spacetime can be described by a single coordinate patch with
a functional form of the metric given by (2.13) just for
$k_{1}=+1$. This patch is somewhat larger than that in Fig. 1
as the upper and lower hyperbolae are located at $UV=4$
({\it i.e.} $r=0$) in the kinkless case. Since this
spacetime is geodesically complete on a single patch it
is possible to foliate its Kruskal diagram with spacelike
hypersurfaces which start from left infinity and cut the
diagram all the way through, to finally reach right
infinity after crossing the horizon twice. This is no
longer the case when a kink is present. The hypersurfaces
would then have to be continued over from the right
infinity of patch $k_{1}=+1$ into a similar surface starting
from left infinity of patch $k_{1}=-1$. Unfortunately, this
continuation is disallowed for all $r\neq M$. Curves of constant
$t$ passing through the two patches offer no solution as they
can only foliate either the entire original region
$I_{+}\cup II_{+}\cup I_{-}\cup II_{-}$ or the entire
new region $III_{+}\cup IV_{+}\cup III_{-}\cup IV_{-}$,
but not simultaneously both. On the other hand, these
curves are not spacelike everywhere as their slope
$dU/dV$ changes sign along each path. Only if the
foliating hypersurfaces are on the extreme hyperbolae
$r=M$ of one patch can they be identified with the
similar hypersurfaces in the other patch. In the Lorentzian
Kruskal diagrams these hyperbolae describe a
Tolman-Hawking wormhole (2.19) which covers the whole domain
of possible values for the scale factor $a$ as one
moves from $t=-\infty$ to $t=+\infty$. The whole of
the wormhole spacetime can then be foliated by a
continuous family of surfaces with $t=T=Const.$

\section{\bf The Quantum Black-Hole Pair}
\setcounter{equation}{0}

In this Section we shall consider the process of thermal
emission in the maximally extended spacetimes of the
two coordinate patches of the
Schwarzschild black hole kink by using the
Green function method [12], thereby checking that, in fact,
the two involved black holes form a pair.
We will see that, quite remarkably,
this does not require recoursing to any complexification
of the physical time $t$. Actually, it will be shown that the
procedure is equally valid and leads to exactly the same
results both when applied to Lorentzian and Euclidean black
holes. The approach used by Hartle and Hawking [12], according
to which one should invoke a general relation between metric
periodicity and gravitational temperature, has always seemed [13]
rather mysterious as it gives no explanation to the use of
the Euclidean version of spacetime. We shall show
that in our model periodicity in the metric follows from a
perfectly justifiable requirement of mathematical completeness,
rather than a suggestive though not very convincing procedure.
This result suggests that a proper evaluation of the thermal
processes in black holes need dealing with what would actually be
the complete black hole system, {\it i.e.} a pair of black
holes, each in a different universe, joined by one wormhole
at their middle internal surface. Considering just one of
such black holes may lead to results which are incomplete
or even paradoxical.

In order to see this, let us first compute the most general
analytic expression for the time $\bar{t}$ entering the
definition of the Lorentzian Kruskal coordinates $U$, $V$.
This will require calculating the function $h(r)$ which is
obtained by integration of (2.5). Using the change of
variables $p=k_{1}\cot\alpha$ and the identification
$\sin\alpha=\sqrt{\frac{M}{r}}$, we finally arrive at
\[\bar{t}=t+4Mk_{1} \left\{k_{1}\cot\alpha -\frac{1}{4}\csc^{2}\alpha
+\ln \left( \left| \left[ \frac{(\sin\alpha-k_{1}\cos\alpha)
\sin^{2}\alpha}{(\sin\alpha+k_{1}\cos\alpha)(2\sin^{2}\alpha -1)}
\right] ^{\frac{1}{2}} \right| \right) \right\}\]
\begin{equation}
+2Mk_{1}k_{2}i(1-k_{2})\pi ,
\end{equation}
where the new sign ambiguity $k_{2}=\pm 1$ arises from the
square root in the argument of the $\ln$, and any real
integration constant has been absorbed in time $t$ which
will now depend on $k_{1}$, $t\equiv t(k_{1})$. The constant
imaginary term in (3.1) is only nonzero for $k_{2}=-1$ and
will prove to be essential for a consistent derivation of
the black hole thermal effects using the Green function
approach. Now, one would again recover (2.6) from (2.7) with
the same requirements as in Section 2 if we re-define the
Kruskal coordinates as follows
\begin{equation}
U=\tilde{U}=\pm k_{2}e^{\frac{\bar{t}_{c}}{4k_{1}M}}
e^{\frac{r}{2M}}(\frac{2M-r}{M})
\end{equation}
\begin{equation}
V=\tilde{V}=\pm k_{1}k_{2}e^{-\frac{\bar{t}_{c}}{4k_{1}M}}
\end{equation}
where
\begin{equation}
\bar{t}_{c}=\bar{t}+4\pi iMk_{1}k_{2},
\end{equation}
with $\bar{t}$ taken to be the real part of (3.1). This choice
leaves expressions for $UV=\tilde{U}\tilde{V}$, $F$, $r$ and
the Kruskal metrics (2.13) and (2.16) real and unchanged. For
$k_{2}=-1$ Eqns. (2.11) and (2.12) become, respectively,
the sign-reversed to (3.2)
and (3.3); {\it i.e} the points $(\bar{t}-4\pi iMk_{1},r,\theta ,\phi)$
on the coordinate patches of Fig. 1 are the points on the new
regions $III_{k_{1}}$ or $IV_{k_{1}}$, on the same figure,
obtained by reflecting in the origins of the respective $U,V$
planes, while keeping metric (2.13)
and the physical time $t$ real and unchanged. This leads to
identifications of hyperbolae in the new, unphysical regions
$III_{k_{1}}$ and $IV_{k_{1}}$ with hyperbolae in,
respectively, original regions $II_{k_{1}}$ and $I_{k_{1}}$.

Let us derive in some detail the thermal effects in our black
hole pair model for the Lorentzian Kruskal metric. The
treatment for the Euclidean case is completely parallel and
essentially leads to the same conclusions.

The evolution of a field along null geodesics as those in Fig.
1 can be described using a quantum propagator. If the field is
scalar with mass $m$, such a propagator will be the one used
by Hartle and Hawking [12] which satisfies the Klein-Gordon
equation
\begin{equation}
(\Box_{x}^{2}-m^{2})G(x',x)=-\delta(x,x').
\end{equation}
We note [12] that for metric (2.13) the propagator $G(x',x)$ will be
analytic on a strip of width $4\pi M$ which precisely is that
is predicted by the imaginary constant component of (3.4), thus
without any need of extending time $t$ into the Euclidean
region. Then, following Hartle and Hawking [12], the amplitude for
detection of a detector sensitive to particles of a given
energy $E$, in regions $I_{+}$ and $II_{-}$, would be
proportional to
\begin{equation}
\Pi_{E}=\int_{-\infty}^{+\infty}d\bar{t}_{c}e^{-iE\bar{t}_{c}}
G(0,\vec{R}';\bar{t}_{c},\vec{R}),
\end{equation}
where $\vec{R}'$ and $\vec{R}$ denote respectively $(r',\theta ',\phi ')$
and $(r,\theta ,\phi)$. Since time $\bar{t}_{c}$ (but not $t$)
already contains the imaginary constant term which is exactly
required for the thermal effects to appear, we need now not
make the physical time $t$ complex. From (3.4) and (3.6) one
can then write
\begin{equation}
\Pi_{E}=e^{4\pi Mk_{1}k_{2}E}\int_{-\infty}^{+\infty}d\bar{t}
e^{-iE\bar{t}}G(0,\vec{R}'; \bar{t}+4\pi iMk_{1}k_{2},\vec{R}).
\end{equation}

Let us next consider a point $x'$ on the hyperbola $r=M$ of
region $II_{+}$, corresponding to patch $k_{1}=+1$. Since
such a hyperbola should be identified with the hyperbola
at $r=M$ of region $I_{-}$ of patch $k_{1}=-1$, the point
$x'$ can be taken to simultaneously belong to the two patches.
Then, one can draw null geodesics starting at $x'$ which
connect such a point with different points $x$ on either
the original region $I_{+}$ of patch $k_{1}=+1$ or the original
region $II_{-}$ of patch $k_{1}=-1$. In the first case, we
obtain from (3.7)
\begin{equation}
P_{a}^{I_{+}}(E)=e^{-8\pi ME}P_{e}^{I_{+}}(E),
\end{equation}
where $P_{a}^{I_{+}}(E)$ denotes the probability for detector to
absorb a particle with positive energy $E$ from region $I_{+}$,
and $P_{e}^{I_{+}}(E)$ accounts for the similar probability
for detector to emit the same energy also to region $I_{+}$,
in the coordinate patch $k_{1}=+1$ corresponding to the
first universe. An observer in the exterior original region
of patch $k_{1}=+1$ will then measure an isotropic background
of thermal radiation with positive energy, at the Hawking
temperature $T_{BH}=(8\pi M)^{-1}$.

For the path connecting $x'$ with a point $x$ on the original
exterior region $II_{-}$ of patch $k_{1}=-1$ in the other
universe, we obtain for an observer in this region,
\begin{equation}
P_{a}^{II_{-}}(-E)=e^{+8\pi ME}P_{e}^{II_{-}}(-E).
\end{equation}
According to (3.9), in the exterior region $II_{-}$ of the
second universe there will appear as well an isotropic
background of thermal radiation, also at the Hawking
temperature $T_{BH}$, which is formed by exactly the
antiparticles to the particles of the thermal bath detected
in region $I_{+}$. Whether or not these two backgrounds are
mutually correlated is an issue which could become of crucial
importance to decide on the problem of the loss of quantum
coherence in black holes [14]. Clearly, if as it seems
most likely they were correlated to each other, one would
then expect quantum coherence to be preserved in the full
process involving simultaneous evaporation of the two
black holes, though relative to observers in each universe,
there would appear to be loss of coherence. This would
be a typical example of an apparent paradoxical result
finding a rather natural explanation when regarded in
its complete framework.

In any event, an evaporation process would then follow
according to which a black hole in a universe will disappear
completely, taking with it all the particles that fell in
to form the black hole and the antiparticles to emit
radiation, by going off through the wormhole of throat
radius $M$ whose other end, which opens up in other
universe, is another black hole which can be regarded to
have been formed from the collapse of massive antifermions,
and also evaporated, giving off exactly the antiparticles
to the particles of the thermal radiation emitted by the
first black hole in the first universe.

For $k_{2}=+1$ we obtain similar hypersurface identifications
as for $k_{2}=-1$. In this case, the identifications comes
about in the situation resulting from simply exchanging the mutual
positions of the original regions $I_{k_{1}}$ and $II_{k_{1}}$
for, respectively, the new regions $III_{k_{1}}$ and
$IV_{k_{1}}$, on the coordinate patches of Fig. 1, while
keeping the sign of coordinates $U$, $V$ unchanged with
respect to those in (2.11) and (2.12); {\it i.e.} the points
$(\bar{t}+4\pi Mik_{1},r,\theta,\phi)$ on the so-modified
regions are the points on the original regions $I_{k_{1}}$
or $II_{k_{1}}$, on the same patches, again obtained by
reflecting in the origins of the respective $U$, $V$ planes,
while keeping metric (2.13) and the physical time $t$
real and unchanged. Thus, expressions for the relations between
probabilities of absorption and emission for $k_{2}=+1$
are obtained by simply replacing region $I_{+}$ for
$IV_{+}$, region $II_{-}$ for $III_{-}$, and energy $E$
for $-E$ in (3.8) and (3.9), so that the same Hawking
temperature $T_{BH}$ is obtained in all the cases. Hence,
"observers" will detect thermal radiation with energy $E<0$
on region $IV_{+}$, and with energy $E>0$ on region $III_{-}$,
in both cases at the Hawking temperature.

By starting with the instantonic Kruskal metric (2.16),
and hence analytically continuing
$\bar{t}\rightarrow i\tilde{\bar{t}}$ in the above analysis,
one can readily see that we attain exactly the same results
as in the Lorentzian treatment. This is not surprising
actually. After all, the thermal emission of black holes
was first discovered using a purely Lorentzian formalism [15].
The set of results achieved so far amounts to the
comments included at the beginning of this Section.
Moreover, these results make quite tempting to conjecture that if
eventually a black hole candidate in our universe is
confirmed to exist, then this would also be a proof that
there is another universe whose particles are exactly the
antiparticles to the particles in ours.

\section{\bf On the Origin of Black-Hole Entropy}
\setcounter{equation}{0}

For the reasons given in the Introduction, it appears now
interesting to calculate an exact expression for the
Euclidean action of a Schwarzschild pair. The issue would
relate with the problem of the origin of the black hole
entropy and, moreover, with the question on whether the
baby universes that nucleate with the black hole pair might
carry some finite entropy. Pair production appears to be
independent of Planck scale physics and therefore it should
be an unambiguous consequence from quantum gravity [7].

The amplitude for production of nonrotating, chargeless
black hole pairs can be calculated in the semiclassical
approximation from the corresponding instanton action
which is
\begin{equation}
S=-\frac{1}{16\pi}\int d^{4}x\sqrt{g}R
-\frac{1}{8\pi}\int d^{3}x\sqrt{h}K,
\end{equation}
where $K$ denotes the trace of the second fundamental form,
and $h$ is the metric on the boundary. We wish to find the
contribution of the two black holes plus their bridging
wormhole to action (4.1). This contribution will correspond
to substracting from (4.1) the flat metrics that the
instanton asymptotically approaches.

Since variation of the tilt angle $\alpha$
must be continuous at $r=M$,
the above contribution can be evaluated from the
corresponding enforced variation
$\delta$ in the time parameter of the instanton at $r=M$
that leads to a net variation of radial coordinate $(\delta r)$
also on $r=M$. This variation would be computed with respect
to a given coordinate patch, as one passes from that patch to
the other, and leaves unchanged the flat metrics. It would
correspond to a variation of action (4.1), $\delta S$, which
should be finite when evaluated at $r=M$, but would diverge
on the event horizon. In what follows we shall interpret
$\delta S|_{r=M}$ as the contribution of the two black holes
and the intermediate wormhole to instanton action (4.1).

Let us then evaluate $\delta S$. Using Einstein equations we find
\[\delta S=-\frac{1}{8\pi}\int d^{3}x\delta(\sqrt{h}h^{ik}K_{ik})\]
\begin{equation}
=-\frac{1}{8\pi}\int d^{3}x\left( \sqrt{h}(\delta h^{ik})(K_{ik}
-\frac{1}{2}h_{ik}K)+\sqrt{h}h^{ik}\delta K_{ik} \right).
\end{equation}
The last term in the rhs of (4.2) can be written
\[h^{ik}\delta K_{ik}=\frac{1}{2}h^{ik}\frac{\partial}{\partial t}\delta h_{ik}
=\frac{1}{2}\frac{\partial}{\partial x^{l}}h^{ik}\dot{x}^{l}\delta h_{ik}
=\frac{1}{2}\frac{\partial A^{l}}{\partial x^{l}},\]
with $A^{l}=h^{ik}\dot{x}^{l}\delta h_{ik}$ a contravariant vector,
and the overhead dot meaning time derivative. Hence
\[h^{ik}\delta K_{ik}=\frac{1}{2}\frac{1}{\sqrt{h}}\frac{\partial}{\partial
x^{l}}(\sqrt{h}A^{l}), \]
so that
\[\int d^{3}x\sqrt{h}h^{ik}\delta K_{ik}
=\frac{1}{2}\int d^{3}x\frac{\partial}{\partial x^{l}}(\sqrt{h}A^{l}).\]
This can be now converted into an integral over $A^{l}$ extended to
the 2-surface surrounding the boundary. Since variations of the field
are all zero on the boundary, this term must vanish and we have
finally
\begin{equation}
\delta S=-\frac{1}{16\pi}\int_{S^{2}}d^{3}x\sqrt{h}
(\delta h^{ik})(2K_{ik}-h_{ik}K),
\end{equation}
where $S^{2}$ denotes the 2-sphere on $r=M$.

For the Schwarzschild metric, we then have
\[\delta S=\frac{M}{\pi}\left[-\frac{1}{24}\rho^{-\frac{3}{2}}
+\rho^{-\frac{1}{2}}+\arctan (\rho^{\frac{1}{2}}) \right]
\int_{0}^{\pi}d\theta\int_{0}^{2\pi}d\phi (\delta r) \]
\begin{equation}
=(-\frac{M}{24}+M+\frac{\pi M}{4})\int_{0}^{2\pi}d\phi (\delta r).
\end{equation}
where $\rho =\frac{2M}{r}-1$.
We evaluate variation $(\delta r)$ at $r=M$ by considering
null geodesics that cross each other at exactly
the surface $r=M$, going always
through original regions on the Kruskal diagrams for the two patches.
For such geodesics we have
$\bar{t}_{k_{1}=+1}=\bar{t}_{k_{1}=-1}$. Thus, if in passing
from the original regions of
patch $k_{1}=+1$ to the original regions of
patch $k_{1}=-1$ time $\bar{t}$
remains constant, time $t$ must change on $r=M$ according
to (Ref. Eqn. (3.1))
\begin{equation}
\left. \delta t \right|_{r=M}\equiv \left. t_{k_{1}=+1}-t_{k_{1}=-1}
\right|_{r=M}=2M.
\end{equation}
In order to calculate the corresponding rate $\frac{dr}{dt}$ from patch
$k_{1}=+1$, we note that
\begin{equation}
\left. \frac{d\bar{t}}{dr} \right|_{r=M}
=\left. (\frac{dt}{dr}+\tan 2\alpha-\frac{1}{\cos 2\alpha})\right|_{r=M}=0.
\end{equation}
Hence,
\begin{equation}
\left. \frac{dt}{dr} \right|_{r=M}=-1.
\end{equation}
We have then
\begin{equation}
(\delta r)_{k_{1}=+1\rightarrow k_{1}=-1}=
\left. \delta t\right|_{r=M} \left. \frac{dr}{dt} \right|_{r=M}=-2M.
\end{equation}
This would be the contribution of just the black hole in the universe
described in coordinate patch $k_{1}=+1$ and the half of the
associated wormhole relative to the same universe. Evaluation of
$(\delta r)_{k_{1}=-1\rightarrow k_{1}=+1}$, corresponding to
the black hole and wormhole half for the other universe, leads
to the same value as in (4.8). Therefore, the full variation
of radial coordinate implied by invariance of the kink number
becomes $(\delta r)=-4M$. Hence, from (4.4) we finally obtain
as the exact expression for the Euclidean action of the two
black holes plus the wormhole in a pair
\begin{equation}
I=\delta S = \frac{\pi}{3}M^{2}-8\pi M^{2}-2\pi^{2}M^{2}.
\end{equation}
The semiclassical production rate is then
\begin{equation}
e^{-I}\sim\exp(-\frac{\pi}{3}M^{2}+8\pi M^{2}+2\pi^{2}M^{2}).
\end{equation}
We interpret now the distinct factors in (4.10). Note first that
the factor $e^{-\frac{\pi}{3}M^{2}}$ should give the full rate
of Schwarzschild black hole pair production in the gravitational
field created by a body with mass $M$ made of fermions in a
universe, and another body with the same
mass, but having exactly the antifermions to the fermions of
the first body, in the other universe.

The second factor in (4.10) gives the sum of the entropies
of the two black holes, each being $S_{BH}=4\pi M^{2}
=\frac{1}{4}A_{BH}$, where $A_{BH}$ is the surface area of
a single black hole. This entropy can also be obtained from
the black hole temperature $T_{BH}$ derived in Section 3 by
insertion into the thermodynamic formula $T_{BH}^{-1}
=\frac{\partial S_{BH}}{\partial E}$.

The last factor in (4.10) should then be interpreted as the
entropy of the nucleated baby universe. This would be a
closed universe with volume $V_{U}=2\pi^{2}R^{3}$, in which
$R$ is the radius of the baby universe, $R=M$. Since the
term $\frac{\pi}{3}M^{2}=S_{p}$ has been interpreted as the
action for the production rate
of a Schwarzschild black hole pair and,
according to (4.9), the total Euclidean action is smaller than
$S_{p}$ by $2S_{BH}$ and $2\pi^{2}M^{2}$, then each of the
factors $e^{2S_{BH}}$ and $e^{2\pi^{2}M^{2}}$ should count
a given number of internal states. One would therefore expect
the latter factor to be associated with a maximum entropy
proportional to the bounded volume of a closed space, that is
\begin{equation}
S_{U}\propto\frac{V_{U}}{l_{p}^{3}}=\frac{2\pi^{2}M^{3}}{l_{p}^{3}},
\end{equation}
where, for a moment, we have explicited the Planck length $l_{p}
=\sqrt{G}$.

However, there is good reason to believe that gravitational
entropy is proportional to the surface area and not the
volume of the bounded region. In this sense, a transition
from volume to area has been proposed by 't Hooft [16] by invoking
a "holographic projection" according to which it is possible
to describe all internal states within a bounded volume by a
set of degrees of freedom which reside on the surface bounding
the given volume, without any loss of
actual information. Because we
now know where the internal states of black holes and wormholes
may reside, a particular implementation of
this volume$\rightarrow$area transition can be given. Thus, if
one would assume that the internal states of a black hole are
in the entire volume $\frac{4}{3}\pi R^{3}$ of a 2-sphere, the
number of such states would be
\[\frac{4}{3}\pi\frac{R^{3}}{l_{p}^{3}}.\]
The actual number of black hole internal degrees
of freedom could then be
estimated by the following ansatz: shrink first any of the
three dimensions to $l_{p}$ and then restrict to the surface
whose degrees of freedom result from holographic projection
of the region
where the states can actually reside. Recalling that black hole
states should all be within the spherical shell between
$2GM$ and $GM$, but not within the spherical region of
radius $r=GM$,
we get for the degrees of freedom whose count would produce
the black hole entropy
\begin{equation}
\left. \frac{4}{3}\pi\frac{R^{2}}{l_{p}^{2}} \right|_{GM}^{2GM}
=4\pi GM^{2}=S_{BH}.
\end{equation}
When applied to a closed space with volume $2\pi^{2}R^{3}$, this
ansatz leads then to
\begin{equation}
S_{U}= \left. 2\pi^{2}\frac{R^{2}}{l_{p}^{2}} \right|_{0}^{GM}
=2\pi^{2}GM^{2}=\frac{1}{2}\pi GA_{U},
\end{equation}
where $A_{U}$ is the surface area for the baby universe. Restoring
full natural units, we see that (4.13) exactly coincides with the last
term in the rhs of (4.9).

If we had evaluated $\delta S$ at the event horizon $r=2M$, instead
of $r=M$, in (4.4), then the action had diverged so that
$\delta S\rightarrow +\infty$, as it was expected. For such
a case, moreover, the rate of pair production had vanished identically,
and the entropy for black holes gone to its classical infinite
value, while that for baby universe vanished. It appears therefore
that black hole pairs joined at the event horizon cannot be
produced as they would correspond to perfectly classical
spacetime constructs.

Consistency of the above interpretations can only be achieved
if we look at the factors $e^{S_{BH}}$ and $e^{S_{U}}$ as
counts of, respectively, either the number of physically
relevant black hole internal
states residing in between the event horizon and the interior
surface at $r=M$, and the number of physically relevant
internal states beyond
$r=M$ of a baby universe, or rather the number of degrees
of freedom on the event horizon surface and on the surface
$R=M$.
Positiveness of the full exponent
in (4.10) leads, on the other hand, to the remarkable feature
that although the rate of pair production is maximum for
Planck-sized black holes, once one of such pairs is formed,
the semiclassical probability (4.10) will tend to favour
processes in which the mass of the black holes, and hence
the size of the baby universe, increase endlessly.

\section{\bf Summary and Conclusions}
\setcounter{equation}{0}

In this paper we have presented arguments in favour of the
existence of an instanton that describes the production of
pairs of chargeless, nonrotating black holes with any mass
$M$, which are joined on the interior surface $r=M$. The
latter feature distinguishes these constructs from the
Einstein-Rosen bridge where junction occurs on the event
horizon surface, at the bifurcation point $U=V=0$ on the
kinkless Kruskal diagram.
We have interpreted these Schwarzschild
black hole pairs as gravitational topological defects that
avoid the curvature singularity at $r=0$.

The geodesic incompleteness at $r=2M$ in the spacetime
metric representing these pairs is
removed by using the Kruskal method, both in the Lorentzian
and Euclidean metrics. The only difference in the
functional form of the resulting two
metrics is in the sign for the metric component for Kruskal
coordinates.

The process of thermal emission by a pair has been studied
by employing the Green function technique in the purely
Lorentzian case, without recoursing to any complex time.
We obtain thus the result that the particles of the
radiation emitted by one black hole are the antiparticles
to the particles in the radiation emitted by the other
black hole. This result suggests that black holes can never
be individually produced, but only in black hole-antiblack
hole pairs, each black hole in a pair being in a different
universe.

An exact expression has been calculated for the action of
the instanton pair. This action is smaller than that for
the corresponding rate of pair production by exactly the
Bekenstein-Hawking entropy of the two black holes,
and by an entropy for baby
universe given by $S_{U}=2\pi^{2}M^{2}$. We conclude
therefore that these entropies count the
number of, respectively, physically relevant
black hole internal states and
baby universe internal states ({\it i.e.} the degrees of
freedom on, respectively, the event horizon and the
interior surface at $r=M$),
and that although the
rate of pair production is maximum for Planck-sized black
holes, there is a small but still nonvanishing probability
for the nucleation of baby universes with macroscopic sizes.
On the other hand, overall positiveness of the exponent of
the semiclassical probability implies that once a Planck-sized
baby universe is created, it will tend to expand, in accordance
with the second law of thermodynamics. From this standpoint,
annihilation of black hole pairs [6] could only occur through
Hawking evaporation or, statistically, in a large ensemble
of pairs while the total entropy of the system still increases.
The model could thus be implemented as a cosmogonic model
as far as it actually implies spontaneous creation
of a closed Tolman universe which can be driven
to expansion. Finally, it is worth noticing that this model
offers a suitable scenario where the
often assumed connection between
black holes and wormholes can be implemented.
Actually, we have seen that
in fact the metric of these wormholes is not but a natural
continuation of the Lorentzian
Schwarzschild half one-kink metric.

\vspace{1.5cm}

\noindent {\bf Acknowledgements}

\noindent This work was supported by CAICYT under Research Project N§
PB91-0052.
The author is indebted to G.A. Mena Marug n for some useful conversations.

\pagebreak

\noindent\section*{References}
\begin{description}
\item [1] G 't Hooft, Nucl. Phys. B256 (1985) 727; V Frolov and I Novikov,
Phys. Rev. D48 (1993) 4545; S Carlip and J Uglum, Phys. Rev. D50 (1994) 2700.
\item [2] V Frolov, {\it Black Hole Entropy}, hep-th/9412211 (1994).
\item [3] D Garfinkle and A Strominger, Phys. Lett. 256B (1991) 146.
\item [4] D Garfinkle, S Giddings and A Strominger, Phys. Rev. D49 (1994) 958.
\item [5] H F Dowker, J Gauntlett, S Giddings and G T Horowitz, Phys.
Rev. D50 (1994) 2662.
\item [6] S W Hawking, G T Horowitz and S F Ross, {\it Entropy, Area,
and Black Hole Pairs}, gr-qc/9409013.
\item [7] G T Horowitz, in: {\it Matters of Gravity}, No. 5 (1995) 10.
\item [8] D Finkelstein and C W Misner, Ann. Phys. (N.Y.) 6 (1959) 230;
D Finkelstein, in: {\it Directions in General Relativity I},
eds. B L Hu, M P Ryan Jr. and C V Vishveshwara (Cambridge Univ. Press,
Cambridge, 1993).
\item [9] A Einstein and N Rosen, Phys. Rev. 48 (1935) 73.
\item [10] S W Hawking, in: {\it Astrophysical Cosmology: Proceedings
of the Study Week on Cosmology and Fundamental Physics}; eds. H A Brck,
G V Coyne and M S Longeir (Pontificiae Academiae Scintiorum Scripta
Varia, Vatican City, 1982), p. 563; J J Halliwell and R Laflamme,
Class. Quant. Grav. 6 (1989) 1839; P F Gonz lez-D¡az, Phys. Rev.
D40 (1989) 4184.
\item [11] D Finkelstein and G McCollum, J. Math. Phys. 16 (1975) 2250.
\item [12] J B Hartle and S W Hawking, Phys. Rev. D31 (1976) 2188.
\item [13] N Pauchapakesan, in: {\it Hightlights in Gravitation and
Cosmology}, eds. B R Iyer, A Kembhavi, J V Narlikar and C V
Vishveshwara (Cambridge Univ. Press, Cambridge, 1988).
\item [14] S W Hawking, Phys. Rev. D14 (1976) 2460; Phys. Rev.
D37 (1988) 904.
\item [15] S W Hawking, Commun. Math. Phys. 43 (1975) 199.
\item [16] G 't Hooft, {\it Dimensional Reduction in Quantum Gravity},
Utrecht Preprint THU-93/26 (1993).

\end{description}

\pagebreak

\noindent {\bf Legend for Figure}

\noindent Fig. 1: Kruskal diagrams for the two coordinate patches
($k_{1}=\pm 1$) of the one-kink extended Schwarzschild metric.
Each of these patches is regarded as a different universe. Points
on the diagrams represent 2-spheres. The null geodesic discussed
in the text is the straight line labelled $a_{1}a_{2}a_{3}$ on the
diagrams. The hyperbolae at $r=M$ are identified on, respectively,
the original regions ($II_{+}$ and $I_{-}$) and the new regions
($III_{+}$ and $IV_{-}$) created by the Kruskal extension.

\end{document}